\documentclass[conference]{IEEEtran}
\usepackage{lipsum} 
\usepackage{adjustbox}
\usepackage{hyperref}
\usepackage[numbers]{natbib}
\usepackage{graphicx}
\usepackage{array}

\begin{document}

\title{Using Language Models for Software Tool Configuration}

\author{\IEEEauthorblockN{Jai Kannan}
\IEEEauthorblockA{\textit{Applied Artificial Intelligence Institute} \\
\textit{Deakin University}\\
Burwood, Australia \\
jai.kannan@deakin.edu.au}
}
\maketitle

\begin{abstract}
In software engineering, the meticulous configuration of software tools is crucial in ensuring optimal performance within intricate systems. However, the complexity inherent in selecting optimal configurations is exacerbated by the high-dimensional search spaces presented in modern applications. Conventional trial-and-error or intuition-driven methods are both inefficient and error-prone, impeding scalability and reproducibility. In this study, we embark on an exploration of leveraging Large-Language Models (LLMs) to streamline the software configuration process. We identify that the task of hyperparameter configuration for machine learning components within intelligent applications is particularly challenging due to the extensive search space and performance-critical nature. Existing methods, including Bayesian optimization, have limitations regarding initial setup, computational cost, and convergence efficiency. Our work presents a novel approach that employs LLMs, such as Chat-GPT, to identify starting conditions and narrow down the search space, improving configuration efficiency. We conducted a series of experiments to investigate the variability of LLM-generated responses, uncovering intriguing findings such as potential response caching and consistent behavior based on domain-specific keywords. Furthermore, our results from hyperparameter optimization experiments reveal the potential of LLMs in expediting initialization processes and optimizing configurations. While our initial insights are promising, they also indicate the need for further in-depth investigations and experiments in this domain.
\end{abstract}


\section{Introduction}
In software engineering, the task of configuring software tools is critical in ensuring optimal performance and functionality of the systems they underpin \cite{zhang2021evolutionary, siegmund2020dimensions}. The configuration process involves the selection and adjustment of parameters and options specific to each usecase \cite{8812912, fahmy2021human}. For example, Think about setting up a tool that checks Python code in a smart app combining software engineering and machine learning. To make it work well, you need to configure it properly to handle both types of code, so it gives better results and smarter insights \cite{kannan2022mlsmellhound}. However, selecting an optimal configuration is complex which is resource and compute-intensive, requiring several iterations.

This complexity is primarily due to the high-dimensional search space \cite{siegmund2020dimensions,blumenschein2023monitoring}. Fully exploring all options is unrealistic \cite{krishna2020cadet,iqbal2022unicorn}, leading engineers to often resort to trial and error or intuition \cite{Tomasdottir2017,Vassallo2020}. However, this method is inefficient and error-prone as it heavily relies on human judgement \cite{Tomasdottir2017, Tomasdottir2020} and impairs the scalability and reproducibility of the process\cite{wood2011impact,hutter2017configurable,bilal2020finding}. Furthermore, the high dimensionality and complexity introduces a challenge where a minor oversight has significant implications on software performance.

The problem of high-dimensional search space for configurations increases as intelligent applications with learning-enabled components have become ubiquitous \cite{langford2023modalas}. Learning-enabled components (LECs) are machine learning components whose behaviour is derived from training data \cite{Casimiro2022} which is integrated into larger systems containing traditional computational entities such as web services and operator interfaces. Despite their widespread use LEC's fail to perform as expected \cite{gesi2023leveraging, langford2023modalas, xiao2022self} which reduces the performance and utility of the system. For instance, a change in a system's operating environment can introduce drifts in the input data for a machine learning (ML) component making them underperform \cite{gesi2023leveraging}.

The optimal performance of an ML component depends on how its hyperparameters are configured during the training process \cite{jin2022deep, shafiq2022deep, hinton2012deep}. This issue becomes particularly pronounced when dealing with extensive search spaces. Sub-optimal parameter settings can lead to noticeable degradation in performance. To address this challenge, a common strategy involves employing Bayesian optimization to systematically explore the search space and identify suitable hyperparameters for retraining or fine-tuning ML components when drifts occur \cite{Casimiro2022}.

The fundamental concept behind Bayesian optimization is to utilize a Gaussian process \cite{rasmussen2003gaussian} to model the intricate relationship between hyperparameter configurations and performance metrics, like validation loss. This model-driven process guides the identification of configurations within the specified search space. However, this approach comes with several limitations. First,  \textbf{i) the process necessitates configuring the starting point or initial conditions for the search space before implementing Bayesian optimization.} Typically, this is achieved through intuition or referencing academic literature, often leading to sub-optimal setups. Secondly, \textbf{ii) if the search space is expansive, the Bayesian optimization process becomes computationally demanding}. Thirdly, \textbf{iii) given a limited budget, the approach may only explore a small fraction of the vast search space, struggling to converge to the optimal configurations}, ultimately resulting in resource wastage \cite{wang2018edge, Standard2019, Ramya2022} and poor performance.

In light of these challenges, we propose that Large-Language Models (LLMs) like Chat-GPT, trained on a diverse array of internet data encompassing machine learning repositories and Python notebooks, can expedite the identification of starting conditions and narrow down the search space for optimal configurations, provided the relevant context. This hypothesis is motivated by the need to address the limitations of existing methods and harness the potential of LLMs to streamline the configuration process.

The expected benefits of exploring LLM driven automated configurations are i) reducing the cost of developing intelligent applications, ii) faster development cycles and iii) improves the quality of LEC's. Ultimately, exploring this research will allow us to enhance the state of software engineering practices.

Our study makes a significant contribution by conducting an initial exploration into the variability of responses generated by ChatGPT-4 across distinct use cases. While our findings offer compelling insights, they also underscore the need for more comprehensive investigations. We identify interesting results, including the possibility of response caching within the Generative Pre-trained Transformer (GPT) models, as well as the diversity observed in responses from RQ2, suggesting consistent and predictable behaviour based on domain-specific keywords. Moreover, our findings from RQ3 indicate the potential of Large-Language Models to expedite hyperparameter initialization and optimization processes. These initial findings not only provide a foundation for further research but also demonstrate the potential of exploring the use of Language Models in software engineering practices.

\label{introduction}


\section{Method}
\subsection{Hypothesis:}
Using Large language models (LLMs) for hyperparameter configuration is a superior technique compared to traditional methods. LLMs have the advantage of understanding context and generating specific configurations that fit the problem. As LLMs have been trained on internet data containing Python notebooks which often contain the results of the experiments within them. LLMs have the edge in mimicking human decision-making processes by leveraging their training data to effectively mimic the decisions of domain experts. We believe we can utilise this feature of the LLMs to bootstrap the configurations for software tools.

This section outlines our method for examining the potential of Large language Models (LLMs) for configuring software components. We employed the Goal-Question-Metric (GQM) framework (Fenton et al., 2014) to define our objective, formulate research questions, and quantify the outcomes of these research questions, facilitating the measurement of goal achievement."

\subsection{Defining the Research Goal:}
The overarching goal of this study is to determine the feasibility of utilising Language Model Models (LLMs) to configure software components automatically. Our aim is to assess the feasibility of utilising Language Model Models (LLMs) for automated software component configuration, focusing on the optimisation of machine learning models to enable intelligent applications. Specifically, we explore whether LLMs can identify optimal starting points for hyperparameter optimisation for an usecase, e.g. \textit{I want to deploy a model on a drone to detect animals in a 30-acre farm}. The objective is to reduce energy consumption and the number of iterations required for fine-tuning a model while adhering to a fixed budget for a particular usecase.

\subsection{Developing Research Questions to Address the Goal:}
The research goal was broken into the following research questions.
\begin{enumerate}
    \item \textbf{\textit{RQ1: What is variability within a specific usecase?}}
   This research question examines potential variations and changes in the distribution of suggested hyperparameters for individual use cases.

    To measure this question, we use statistical measures of standard deviation, variance and interquartile ranges to calculate the variability.

    The investigation allows us to study the performance diversity of a single usecase.
   \item \textbf{\textit{RQ2: How does the variability differ across multiple usecases?}}
    This research question explores how variability in LLM-generated hyperparameters differs across diverse use cases.

    To measure this research question, we used statistical analysis using ANOVA's F-statistic measure to compare the generated hyperparameter across diverse usecases.

   This investigation helps us understand the adaptability and flexibility of LLMs in meeting distinct requirements within each application domain.

    \item \textbf{\textit{RQ3: How does the performance of LLM-configured software tools compare with state-of-the-art methods?}}
    This research question assesses the competitive advantage of LLMs in configuring software tools, comparing their performance against State-Of-The-Art (SOTA) methods, such as Bayesian optimization.

    To measure this research question, we calculate the performance of the tuned models using the two approaches i.e. LLM-based and Bayesian optimisation on a dataset, and calculate the accuracy can compare the validation losses for each approach.

    This analysis enables us to identify whether LLMs can accurately predict discrete values.
\end{enumerate}



\label{method}

\section{Experiment Setup}

To evaluate the research questions, selecting an LLM becomes crucial. LLMs represent a significant advancement in artificial intelligence, comprehending and generating human-like text from the input. Their applications span across diverse domains such as natural language processing, content generation, and text summarizations. Several LLM options available for consideration such as LLaMA, LLaMA2\footnote{https://about.fb.com/news/2023/07/llama-2} from Meta, BLOOM\footnote{https://huggingface.co/docs/transformers/modeldoc/bloom} from Hugging Face, and ChatGPT \footnote{https://openai.com/gpt-4} from Open-AI. For our experiment, we chose to use ChatGPT-4. Unlike other models, Chat is specifically tailored for conversational interaction, allowing it to produce coherent and contextually relevant responses. ChatGPT is trained on a diverse range of internet data to generate content for various subjects. This specialisation is crucial in applications requiring interactive dialogue. In this experiment, we use this feature to generate the usecases for answering the research questions.


For the experiments, we chose the domain of computer vision. The popularity of computer vision has seen many real-world applications and opportunities. we chose two usecases in this domain for our experiments, which are described in the following section. Each usecase is chosen due to their tradeoffs while choosing and selecting the Machine learning technique to apply in terms of resources and performance.
 
\subsection{Usecase 1:} \textbf{\textit{Real-time image classification for security cameras:}}
In security applications, real-time image classification is essential for the identification of potential threats. The model should provide rapid predictions for quick decision-making. The model should be compute efficient to execute in a resource constraint environment to ensure real-time processing without compromising image quality.
\subsection{Usecase 2:} \textbf{\textit{Understanding financial market structures for investments:}}
In the case of financial markets, unlike usecase 1, the application is focused on improving the accuracy of diagnosing markets using financial news articles to make informed investment decisions. 

Comparing these scenarios underscores their distinct nature. The security use case prioritises speed, even at a minor expense in accuracy, while the usecase of understanding the financial markets demands precision, allowing for slightly extended inference times to ensure accurate assessments. These use cases exemplify the variability in hyperparameter configuration driven by the specific requirements of each situation along with the model used.

\subsection{Prompt design for usecases:}
To interact with Chat-GPT, we utilise a prompt. A prompt is a specific input query that is provided to the model to receive a response. It is used to initiate a conversation to request information from the model. Chat-GPT utilises two types of prompts i.e. i)System prompt and ii) User prompt, which are described as follows:
\begin{itemize}

    \item \textbf{System prompt:} The system prompt is an initial instruction that is provided to set the context for the conversation. The prompt provides the background information and guidelines for a conversation
    \item  \textbf{User prompt:} The user prompt is the input provided to continue the conversation after the system prompt to query the model for information.
\end{itemize}

\begin{figure}[h]
    \centering
    \includegraphics[width=\linewidth, height=4cm]{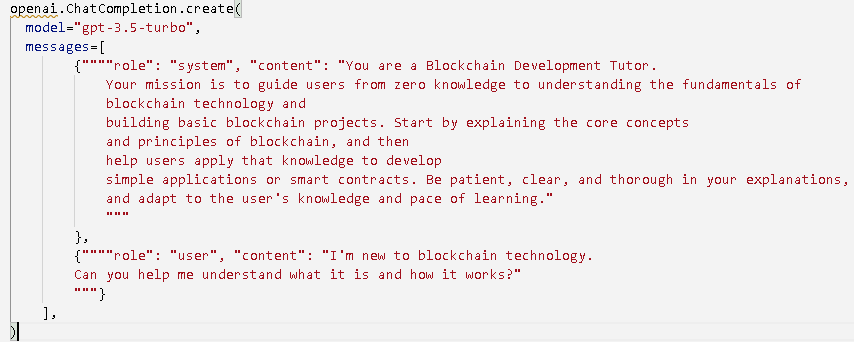}
    \caption{Example demonstrating system and user prompts to set the role of the Chat-GPT model}
    \label{fig:example_sys_user}
\end{figure}

These prompts work together to create a relevant conversation with the model. The system prompt provides the initial context, and the user prompts guide the ongoing conversations. An example of the prompts is presented in \autoref{fig:example_sys_user}
Using a combination of user and system prompts, we can guide the GPT-4 model to provide responses that align with the intended context and purpose of the conversation. The system prompt serves as a crucial foundation, outlining the overall direction and scope of the conversation. It provides the initial context, ensuring that the model understands the topic, tone, and objectives of the interaction.

Once the context is set by the system prompt, the user prompts come into play to lead the conversation forward. These prompts act as queries or instructions that steer the discussion in specific directions. By carefully constructing user prompts, we can extract the desired information, insights, or responses from the model. 
\subsection{Prompting strategies:}

For our experiments, we utilised the combination of 2 different strategies to develop the prompts for each usecase. We utilised the two different strategies as described in \cite{kaddour2023challenges} which are \textbf{1.Instruction-Following} and \textbf{2. Imitation }. The combination of the two strategies allows the crafting of prompts that elicit meaningful and contextually relevant responses from the Chat-GPT model.
\begin{enumerate}
    \item \textbf{Instruction-Following:} Instruction following described by \cite{kaddour2023challenges} is a strategy where prompts are constructed as clear and explicit instructions that guide the model's response. These instructions lay out the desired format, content, or steps the model should follow when generating its reply. This approach is particularly useful when precise and specific answers are needed. 
    \item \textbf{Imitation:} This strategic approach uses a deliberate sequence of queries intended to imitate an interactive discourse with a subject expert. The primary objective of this strategy is to replicate the process of seeking clarification and validation typically found in human-machine interaction.The imitation seeks to replicate the dynamics of a consultation, where the user assumes the role of the inquirer seeking expert insights. 
\end{enumerate}

\subsection{Prompts using Imitation with Instruction Following:}
For designing the prompt using Instruction-following, we designed a prompt structure with the following combination of a system prompt setting the context to imitate a subject-matter expert and user prompts to then query the model. The elements of the prompt, are shown in \autoref{fig:instf_strat}:
\begin{figure}[h]
    \centering
    \includegraphics[width=\linewidth]{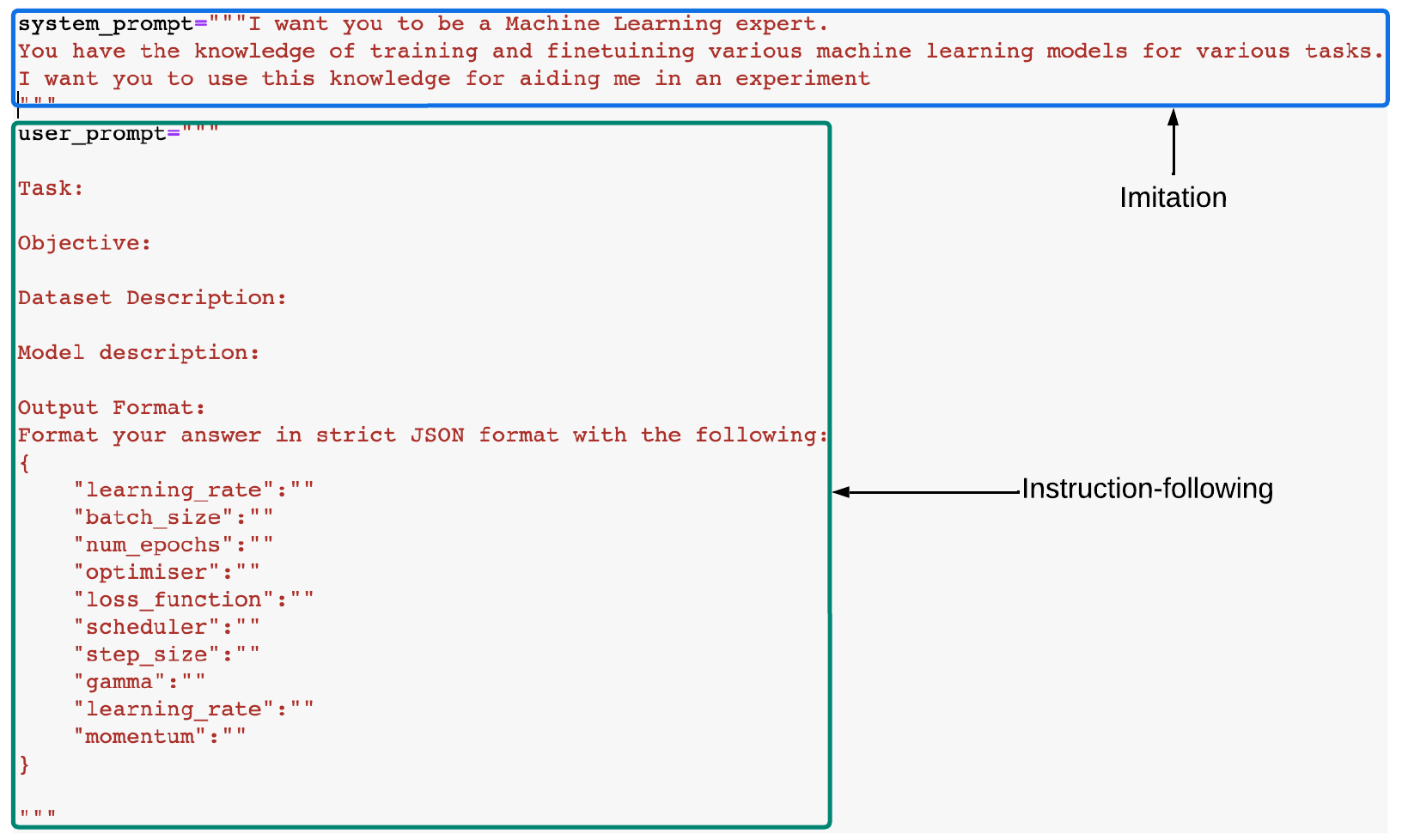}
    \caption{Example of the prompt structure for imitation with instruction-following strategy}
    \label{fig:instf_strat}
\end{figure}

\begin{itemize}
    \item \textbf{System Prompt:} This prompt sets the initial context for our experiment. In our case, we wanted the LLM to imitate a machine learning expert, who has conducted numerous experiments and possesses the knowledge from those experiments for our usecases. We used the following system prompt to set the context for the conversation, \textit{"I want you to be a Machine Learning expert. You have the knowledge of training and finetuining various machine learning models for various tasks. I want you to use this knowledge to aid me in an experiment"}
    \item \textbf{Task:} This describes the core task which need to be completed or achieved.
    \item \textbf{Objective:} States what the user wants to achieve with the task description or the end goal.
    \item \textbf{Dataset Description:} Provides a brief description of the dataset used for the task. This can be obtained from the dataset source.
    \item  \textbf{Model Description:} Provides a brief description of the model being used for the experiment obtained from the model cards.
    \item \textbf{Output format:} Describes how the used wants to generate the output from the LLM.
\end{itemize}
\subsection{Prompts for each Usecase:}\label{sec:usecases}

\begin{figure}[h]
    \centering
    \includegraphics[width=\linewidth]{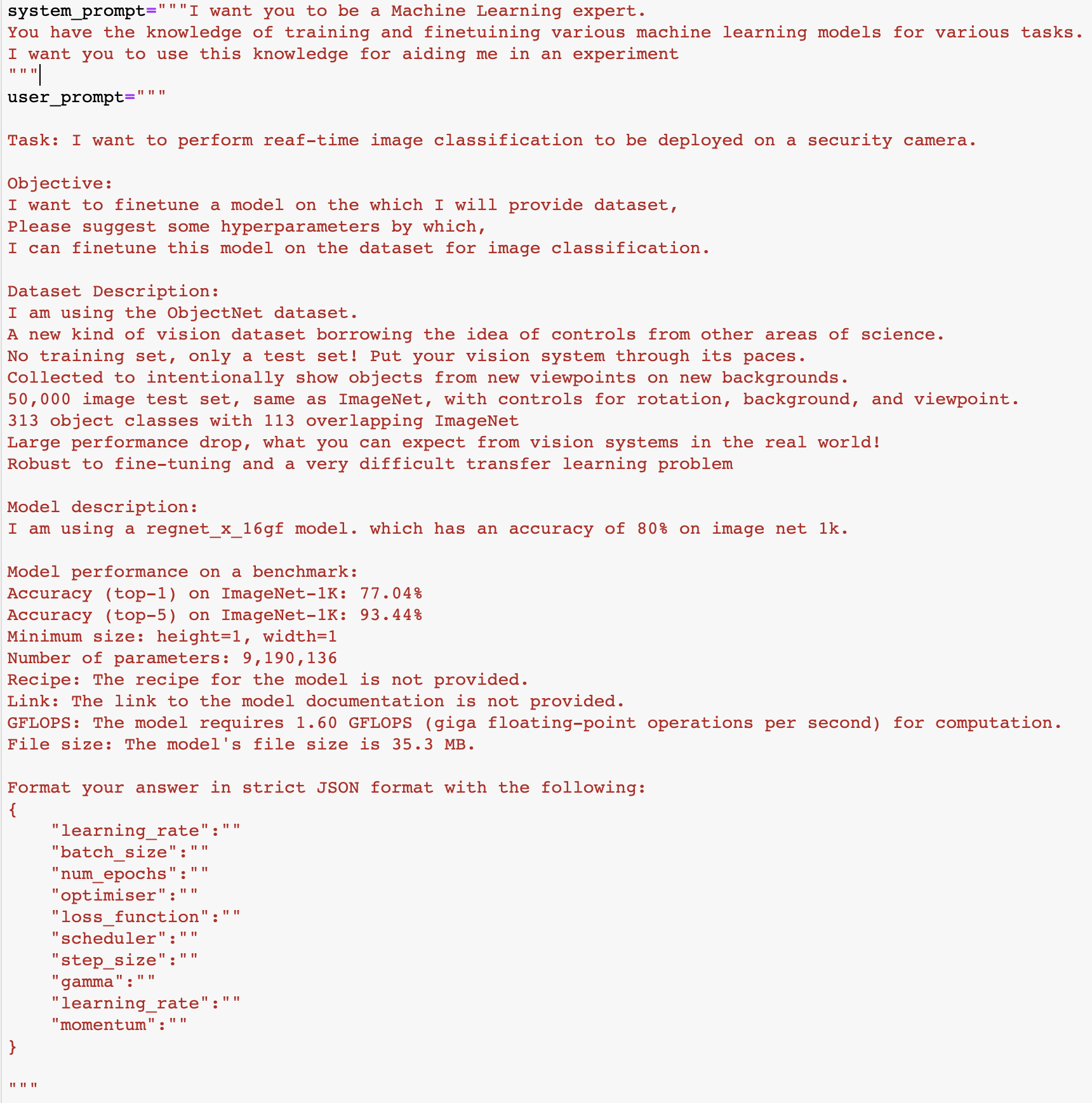}
    \caption{Prompt used for usecase 1 \textit{Real-time image classification for security cameras.}}
    \label{fig:usecase1ifp}
\end{figure}

Using imitation with instruction-following strategy, we developed the following prompt for \textbf{\textit{Real-time image classification for security cameras}} where we decided to use the ObjectNet 2019 \cite{barbu2019objectnet} containing images with unconventional views, contextual variations, and ambiguous scenarios as the dataset. For the model, we decided to use an off-the-shelf Regnet model from Pytorch model repository. Using the Model card, we provided the description for the model being used as well. For the objective, we stated that we were performing fine-tuning on a pre-trained model. Finally, we stated that we wanted the hyperparameters to be outputted in a JSON format for conducting our experiment. \autoref{fig:usecase1ifp} shows the complete prompt for usecase1.

\begin{figure}[h]
    \centering
    \includegraphics[width=\linewidth]{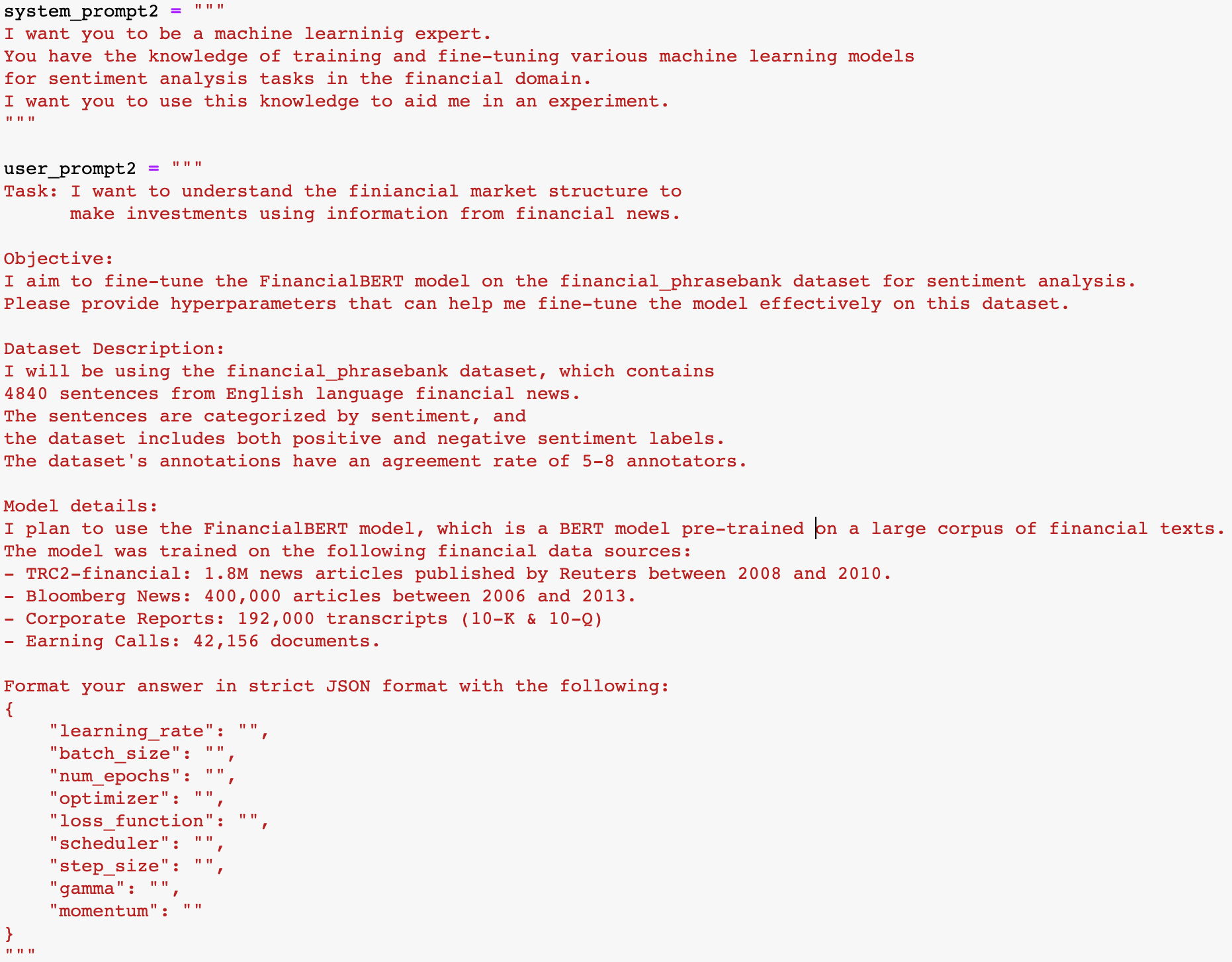}
    \caption{Prompt for usecase 2 \textit{Understanding financial market structures for investments.}}
    \label{fig:usecase2}
\end{figure}

Similar to our approach for usecase 1 using the same imitation with instruction-following strategy, we developed the prompt for \textbf{\textit{understanding financial market structures for investments}}. We decided to use the financial-phrasebank dataset \cite{Malo2014GoodDO} containing 4840 sentences from English language financial news categorised by sentiment. We decided to use an off-the-shelf FinancialBERT \cite{hazourli2022financialbert} model from the Hugging Face model repository. We stated our objective was to fine-tune this model on our dataset and queried the model to suggest hyperparameters for this experiment in JSON format. \autoref{fig:usecase2} shows the complete prompt used for this experiment.
\subsection{Design for research questions 1 and 2:}
To address research questions 1 and 2, we adopted the study design outlined in \autoref{fig:rq1,2}. The prompts generated in \autoref{sec:usecases} were employed to query the GPT-4 model. For our experiments, we set the model's temperature to 0, effectively curbing its inherent creativity. This was done to mitigate the potential generation of content that deviates from the provided source, due to issues in encoding and decoding between text and representations as discussed by Kaddour et al. (2023). By choosing a temperature of 0, we aimed to render the responses predominantly deterministic, while allowing a slight degree of residual variability\footnote{https://platform.openai.com/docs/models/overview}.

\begin{figure}[h]
    \centering
    \includegraphics[width=\linewidth]{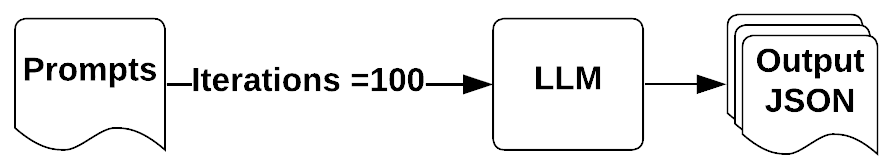}
    \caption{Design for research questions 1 and 2.}
    \label{fig:rq1,2}
\end{figure}

The design depicted in \autoref{fig:rq1,2} guided our methodology. Under these specifications, we executed the prompts for each use case over a span of 100 iterations. We examined the GPT-4 model's responses, to answer research questions 1 and 2. JSON-formatted responses were gathered from the model, providing a snapshot of hyperparameter configurations for each use case. This enabled us to ascertain both the extent of response variability (RQ1) within each use case and the divergence in prompts across different use cases (RQ2).

To measure the significance of our findings, we employed appropriate statistical tests to our research questions. For assessing response variability within each use case (RQ1), we computed variance or standard deviation for responses, comparing these metrics across use cases. For prompt diversity between use cases (RQ2), we calculated Jaccard index between the two usecases. We applied ANOVA or Kruskal-Wallis tests to determine if these metrics differed significantly between use cases. 


\subsection{Design for research question 3:}\label{sec:rq3}
To investigate research question 3, we wanted to identify if an LLM can provide a sensible search space for finetuning hyperparameters as compared to approaches such as Bayesian optimisation. We used a computer vision as an exemplar to demonstrate if LLMs can provide a better search space for configuring hyperparameters. We decided to go with the Regnet model from usecase 1. The model is trained on the ImageNet 1k dataset. The ImageNet dataset is a widely used collection of labelled images covering thousands of object categories, serving as a foundational resource for training and evaluating computer vision models.

In our study, we utilised a filtered version of the ObjectNet dataset. This choice was driven by two key factors: first, the dataset encompasses 113 classes that match those in ImageNet; and second, it rigorously assesses object recognition models against real-world complexities, such as atypical angles and diverse backgrounds, with the intention of bolstering their resilience in real-world situations. Our approach involved extracting images from ObjectNet that corresponded to ImageNet's class categories, and subsequently assigning ImageNet's class labels to these ObjectNet-derived images. We collected 2000 images from ObjectNet with the filtering criteria applied to the images as the dataset for our experiment. We did this to fine-tune the pretrained Regnet model from Pytorch model repository has already been trained using these labels from ImageNet to fastrack our experiments using a 70-30 split of the dataset containing 2000 images. Furthermore, for each experiment we froze all the layers of the model except the last layer on which the training and fine-tuning was performed.  

We set up four experiments with the following experiments \textbf{1) fine-tuning using standard Bayesian optimisation, 2) fine-tuning using hyperparameters from  \cite{kumar2022fine,goyal2022vision} research for Bayesian optimisation 3) fine-tuning using LLM to configure the search space for Bayesian optimisation for hyperparameter optimisation and 4) Using the results from the experiment iii to query an LLM to suggest starting a smaller search space for the Bayesian optimisation of hyperparameters}. In the following subsections, we describe the setup for each of the experiments.

\subsubsection{ Fine-tuning using standard Bayesian optimisation:}
In this experiment, we adopted the standard Bayesian optimisation approach to fine-tune the pretrained Regnet model. We utilised the Hyperopt\footnote{https://github.com/hyperopt/hyperopt} library to initialise the optimisation process. 

Hyperopt uses a search space which defines the range of potential values of hyperparameters that are explored during the optimisation process. This allows the exploration of various options for each hyperparameter, allowing the optimisation algorithm to evaluate and determine the beast configuration for a machine learning model.  The selection of an appropriate search space is critical, as it directly influences the effectiveness and efficiency of the hyperparameter tuning process.

To execute this experiment, we use the following configuration for the search space presented in \autoref{table:config1}.

\begin{table}[h]
\centering
\caption{Hyperparameter configuration for Fine-tuning using standard Bayesian optimisation with random initialised search space}
\begin{tabular}{|>{\hspace{0pt}}m{0.438\linewidth}|>{\hspace{0pt}}m{0.429\linewidth}|}
\hline
Parameters & Value Ranges \\ 
\hline
Learning Rate & 1e-5 , 1e5 \\ 
\hline
Momentum & 0.0 , 1.0 \\ 
\hline
Batch Size & 32 \\ 
\hline
No. of Epochs & 3 \\ 
\hline
No. of Trials & 10 \\ 
\hline
Gamma & 1e-5 , 1e5 \\ 
\hline
Step Size & 0 , 20 \\
\hline
\end{tabular}\label{table:config1}
\end{table}

We executed the optimisation process with the configuration in \autoref{table:config1} for 10 trials containing 3 epochs each due to the resource restrictions imposed by a single threaded CPU execution.

\subsubsection{ Fine-tuning using hyperparameters from research papers for Bayesian optimisation:}
Using research papers to find hyperparameters is a common approach for fine-tuning machine learning models. For this experiment, we referred to the research papers by \cite{kumar2022fine} and \cite{goyal2022vision}. We utilised these papers as they describe the parameters used for fine-tuning the Regnet models, same as the model we are using for our experiment. In \autoref{table:paperconfig} we describe the configuration for fine-tuning with the configurations obtained from the research papers \cite{kumar2022fine, goyal2022vision} for the experiment.
\begin{table}[h]
\centering
\caption{Hyperparameter configuration for Fine-tuning using LLM recommended search space for Bayesian optimisation}
\begin{tabular}{|>{\hspace{0pt}}m{0.438\linewidth}|>{\hspace{0pt}}m{0.429\linewidth}|} 
\hline
Parameters & Value Ranges \\ 
\hline
Learning Rate & 0.9 \\ 
\hline
Momentum & 0.015 \\ 
\hline
Batch Size & 32 \\ 
\hline
No. of Epochs & 3 \\ 
\hline
No. of Trials & 10 \\ 
\hline
Gamma & 0.1 \\ 
\hline
Step Size & 8, 12 \\
\hline
\end{tabular}\label{table:paperconfig}
\end{table}

\subsubsection{Fine-tuning using LLM to configure the search space for Bayesian optimisation for hyperparameter optimisation:}
In this experiment, we queried an LLM to suggest the starting conditions for Bayesian optimisation. We provided the LLM with a prompt similar to the prompt described in \autoref{sec:usecases} where we described the task and objective utilising the imitation with instruction-following strategy. The exception in this case is that we queried the search space initialisation using the ChatGPT UI. The following \autoref{table:LLmsuggestedconfig} shows the search space configurations suggested by the GPT-4 Model.

\begin{table}[h]
\centering
\caption{Hyperparameter configuration for Fine-tuning using LLM recommended search space for Bayesian optimisation}
\begin{tabular}{|>{\hspace{0pt}}m{0.438\linewidth}|>{\hspace{0pt}}m{0.429\linewidth}|} 
\hline
Parameters & Value Ranges \\ 
\hline
Learning Rate & -4 , -2 \\ 
\hline
Momentum & 0.001 , 0.01 \\ 
\hline
Batch Size & 32 \\ 
\hline
No. of Epochs & 3 \\ 
\hline
No. of Trials & 10 \\ 
\hline
Gamma & -8 , -3 \\ 
\hline
Step Size & 10 , 20 ,30 \\
\hline
\end{tabular}\label{table:LLmsuggestedconfig}
\end{table}
\subsubsection{Using the results from the experiment iii to query an LLM to suggest starting a smaller search space for the Bayesian optimisation.}
In our fourth experiment, we aimed to explore whether an LLM could leverage prior knowledge of fine-tuning or training procedures to identify enhanced search spaces for hyperparameter configuration in model fine-tuning. For this experiment, we utilised the results i.e. the validation loss and the validation accuracy  from the previous experiment along with the hyperparameters used to query an LLM to identify a new set of hyperparameters by which we can improve the fine-tuning process by reducing the no. of trials required to achieve the performance from the previous experiment. The configuration used for this experiment is present in the \autoref{table:LLmfinetuned}.

\begin{table}[h]
\centering
\caption{Hyperparameter configuration for Fine-tuning using prior configurations and performance}
\begin{tabular}{|>{\hspace{0pt}}m{0.438\linewidth}|>{\hspace{0pt}}m{0.429\linewidth}|} 
\hline
Parameters & Value Ranges \\ 
\hline
Learning Rate & 0.01 , 0.03 \\ 
\hline
Momentum & 0.001 , 0.01 \\ 
\hline
Batch Size & 32 \\ 
\hline
No. of Epochs & 3 \\ 
\hline
No. of Trials & 10 \\ 
\hline
Gamma & 0.0001 , 0.001 \\ 
\hline
Step Size & 15 , 25 \\
\hline
\end{tabular}\label{table:LLmfinetuned}
\end{table}

\subsubsection{Metrics collection:}
To measure the significance of our findings, we employed appropriate statistical tests tailored to our research questions. For assessing response variability within each usecase (RQ1), we computed variance or standard deviation for responses, comparing these metrics across use cases. For prompt diversity between use cases (RQ2), we applied ANOVA tests to determine if these metrics differed significantly between use cases.

For (RQ3) we calculated the validation loss and validation accuracy for each of the experiments described in \autoref{sec:rq3}. We visualised the results for all the experiments and reported our findings in \autoref{results}.

\label{setup}

\section{Experimental Results}
In this section, we present the outcomes of our experiments aimed at addressing the research questions outlined in the previous sections. The results are organised according to each research question, highlighting the key findings.

\subsection{\textbf{RQ1: What is variability within a specific usecase?}}
To investigate the variability within responses generated by the ChatGPT-4 model, we analysed the JSON-formatted responses obtained from 100 iterations of each use case.  

To investigate the variability within specific use cases, we conducted a thorough analysis of selected columns pertaining to different attributes. The subsequent sections present the outcomes of our statistical tests, highlighting the standard deviations and variances observed within each use case.
\subsubsection{Usecase 1: Real-time image classification for security cameras.}

For Usecase 1, we examined the standard deviations and variances of the following parameters \texttt{'Learning Rate'}, \texttt{'Momentum'}, \texttt{'Batch Size'}, \texttt{'Num Epochs'}, \texttt{'Step Size'}, and \texttt{'Gamma'}. The results indicate a negligible variability within these attributes. Specifically:

\begin{itemize}
  \item \textbf{Learning Rate}: The standard deviation was found to be approximately $6.505 \times 10^{-19}$, with a variance of about $4.232 \times 10^{-37}$.
  \item \textbf{Momentum}: Similarly, the standard deviation was close to $7.772 \times 10^{-16}$, accompanied by a variance of roughly $6.040 \times 10^{-31}$.
  \item \textbf{Batch Size, Num Epochs, and Step Size}: These attributes exhibited a standard deviation and variance of 0, suggesting a lack of variability.
  \item \textbf{Gamma}: The standard deviation was approximately $1.943 \times 10^{-16}$, with a variance of about $3.775 \times 10^{-32}$.
\end{itemize}

\subsubsection{Usecase2: Understanding financial market structures for investments.}
Similar to the previous usecase we conducted the same statistical tests to calculate the standard deviations and variances for the usecase. Similar to usecase 1 we examined the following parameters \texttt{'Learning Rate'}, \texttt{'Momentum'}, \texttt{'Batch Size'}, \texttt{'Num Epochs'}, \texttt{'Step Size'}, and \texttt{'Gamma'}.
\begin{itemize}
  \item \textbf{Learning Rate}: The standard deviation was calculated to be approximately $3.727 \times 10^{-20}$, with a corresponding variance of about $1.389 \times 10^{-39}$.
  \item \textbf{Momentum}: Akin to the previous use case, the standard deviation was close to $7.772 \times 10^{-16}$, accompanied by a variance of roughly $6.040 \times 10^{-31}$.
  \item \textbf{Batch Size, Num Epochs, and Step Size}: These attributes consistently exhibited a standard deviation and variance of 0, indicative of low variability.
  \item \textbf{Gamma}: The standard deviation aligned closely with the previous use case, at approximately $1.943 \times 10^{-16}$, with a variance of about $3.775 \times 10^{-32}$.
\end{itemize}

The consistent trend of low standard deviations and variances across both use cases displays a pattern of minimal variability within the selected attributes. These findings suggest that the examined attributes demonstrate a high degree of consistency and lack of deviation from their respective means. 
\subsection{Findings from RQ1:}
In this section we discuss the findings from our experiments and draw possible conclusions.
\subsubsection{Lack of Variance and Potential Explanations}

The consistent pattern of minimal variability observed within the selected attributes across both use cases prompts us to explore potential explanations for this phenomenon. 
One plausible explanation for the consistently low variability is the possibility of response caching within the Generative Pre-trained Transformer (GPT) models. It is conceivable that the model may have learned and cached certain responses, leading to a uniformity of outcomes for specific prompts. This behaviour is further reinforced by the fact that setting the temperature parameter to 0, effectively reducing randomness during text generation, appears to accentuate this effect. The use of a temperature parameter controls the level of randomness in the generated output. A value of 0 enforces a deterministic response, which could align with the observed uniformity.
\subsubsection{Impact of keywords}
Furthermore, We tried to reword and rephrase the prompts in an attempt to introduce diversity into the generated responses. Despite these efforts, the Language Model (LM) consistently provided similar responses when presented with prompts related to computer vision across different use cases. This finding is intriguing as it suggests a degree of consistency and predictability in the LM's responses for specific domains or topics or the use of specific keywords such as "Classification", "Fine-tuning" or a description of a particular task.

\subsection{\textbf{RQ2: How does the variability differ across multiple usecases?}}
The investigation into variability across multiple usecases was carried out through an Analysis of Variance (ANOVA) test, probing the differences in variability among selected attributes. This section presents the outcomes of the ANOVA test:
\begin{itemize}
  \item \textbf{Attribute: Learning Rate}\\
  The ANOVA analysis for the 'Learning Rate' attribute produced a remarkably high F-statistic, implying a substantial variation in variability across different usecases. The corresponding p-value of 0.0 indicates a significant departure from the null hypothesis, signifying notable differences in the dispersion of data points among the usecases.


  \item \textbf{Attribute: Batch Size}\\
  In parallel to the 'Learning Rate', the ANOVA test applied to the 'Batch Size' attribute demonstrated a remarkably elevated F-statistic and a p-value of 0.0. These results identify substantial differences in variability across usecases with respect to the 'Batch Size'.

  \item \textbf{Attribute: Num Epochs}\\
  Analogously, the 'Num Epochs' attribute revealed significant variability differences across usecases, as indicated by the exceedingly high F-statistic and the associated p-value of 0.0. 

  \item \textbf{Attribute: Step Size}\\
  Further reinforcing the theme of variability disparities across usecases, the ANOVA analysis for the 'Step Size' attribute yielded a conspicuously high F-statistic and a p-value of 0.0. This underscores the importance of accounting for diverse requirements in terms of step sizes for different application contexts.

\end{itemize}

The results from the ANOVA tests identify variability across multiple usecases. Attributes such as 'Learning Rate', 'Batch Size', 'Num Epochs', and 'Step Size' display pronounced differences in variability, signifying the distinct contextual demands associated with different application scenarios

\subsection{Findings from RQ2:}
In this section we discuss our findings from RQ2.

The investigation into variability across multiple usecases has revealed how Language Model (LM) discerns differences between various application scenarios. The analysis of selected attributes has showcased substantial variations in variability, shedding light on the LM's capacity to adapt its responses in accordance with specific usecases.

Notably, attributes such as 'Learning Rate', 'Batch Size', 'Num Epochs', and 'Step Size' exhibited differences in variability across different usecases. 
\subsubsection{Observations Regarding 'Momentum' and 'Gamma'}
An interesting observation concerning the 'Momentum' and 'Gamma' attributes. It is to be noted that these attributes demonstrated consistent variability patterns across both usecases. This uniformity prompts us to consider a compelling hypothesis: the LM might have learned these specific parameter values as commonly used defaults in notebooks and paper implementations. 

\subsection{\textbf{RQ3: How does the performance of LLM-configured software tools compare with state-of-the-art methods?}}
\begin{figure}[h]
    \centering
    \includegraphics[width=\linewidth]{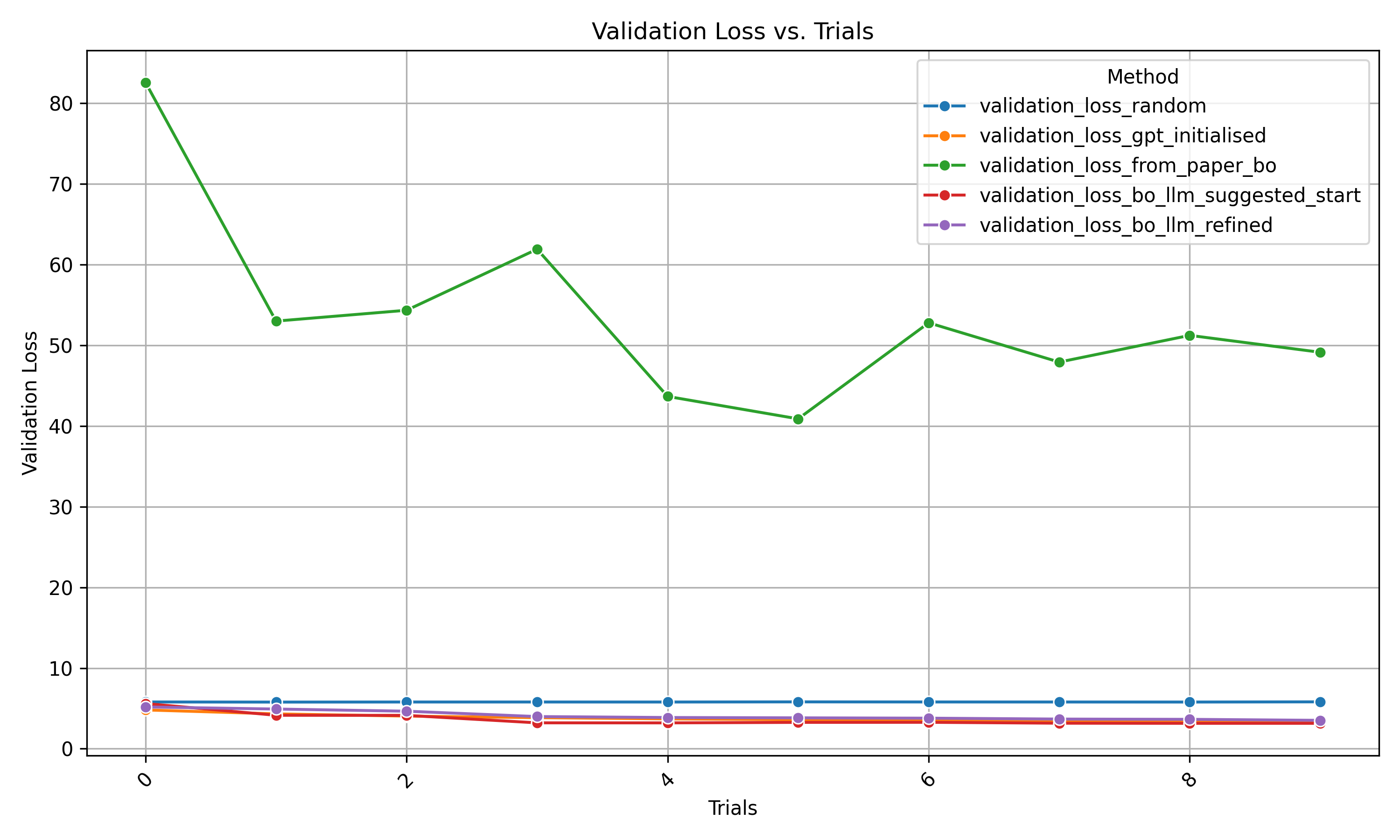}
    \caption{Validation losses for each experiment}
    \label{fig:val_loss}
\end{figure}

\begin{figure}[h]
    \centering
    \includegraphics[width=\linewidth]{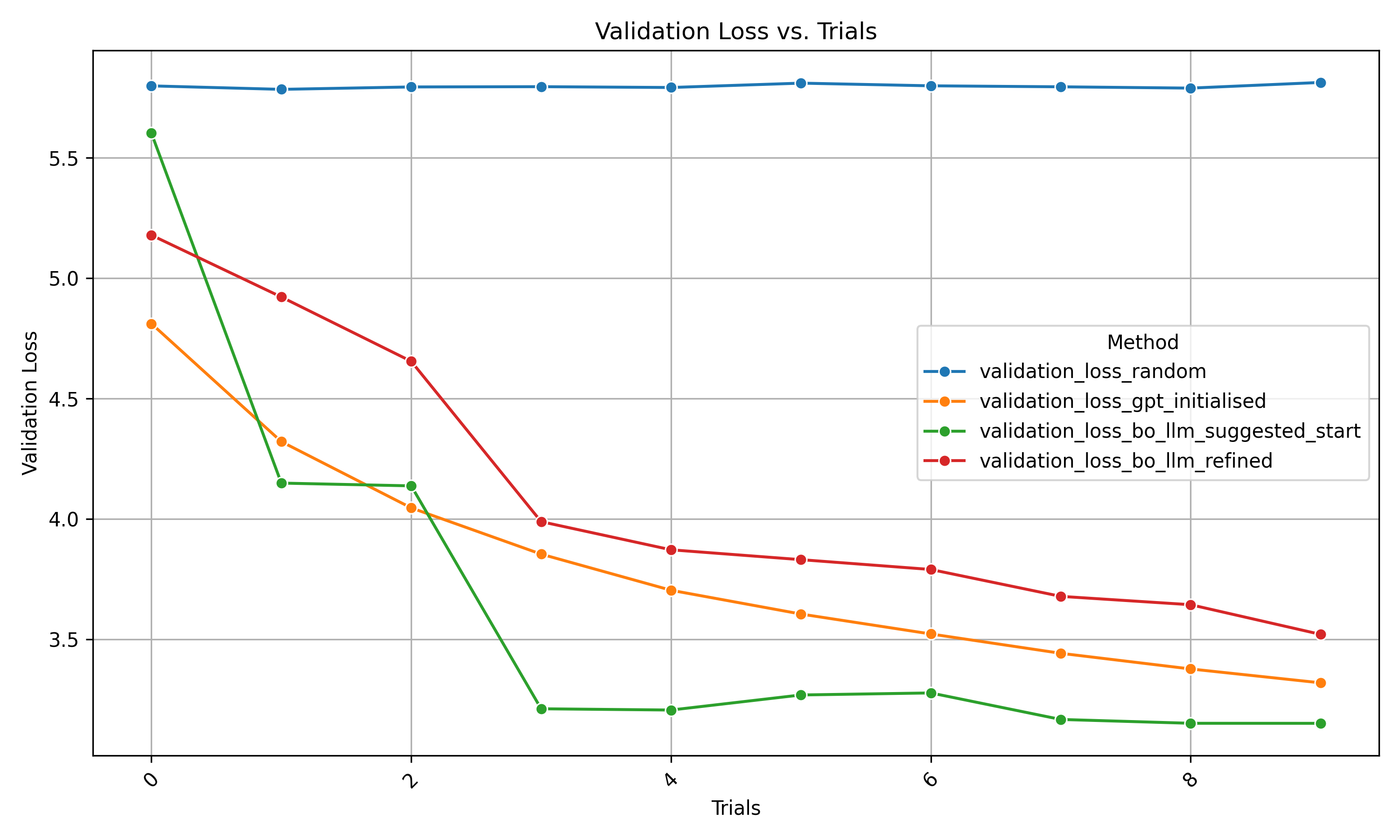}
    \caption{Validation losses after dropping the configurations obtained from the research papers.}
    \label{fig:val_loss2}
\end{figure}

For this experiment we ran for experiments with different hyperparameter configurations. For the first set of experiments outlined in \autoref{sec:rq3} we compared the the validation losses initialising the i)Bayesian optimisation process with a random start, ii)parameters obtained from a research paper, iii) using an LLM to suggest the start conditions and search space for the optimisation process and finally using the outputs collected from step 3 to inform, finding a new search space using an LLM. The results of the experiment are displayed in \autoref{fig:val_loss}.
Analysis of Figure \autoref{fig:val_loss} reveals distinct trends among the hyperparameter configurations. Notably, the configuration stemming from research papers exhibits the highest validation loss, while the remaining configurations closely align with each other in terms of validation loss during the fine-tuning process. To delve deeper into the fine-tuning process, we excluded the results from the research paper configurations and examined the results in Figure \autoref{fig:val_loss2}. 

Figure \autoref{fig:val_loss2} identifies interesting observations. In an additional experiment, we inquired the LLM for hyperparameters and fine-tuned the model following the prompt structure outlined in Figure \autoref{fig:usecase1ifp}, without Bayesian optimization. This inquiry is represented by the yellow line in the graph. Furthermore, it becomes evident that when utilizing the LLM to derive initial conditions and search spaces for Bayesian optimization, as indicated by the green line in Figure \autoref{fig:val_loss2}, a more rapid reduction in validation loss is observed, indicative of a better model convergence.
\subsubsection{Findings from RQ3}
In this section we discuss our findings from RQ3.

In the context of machine learning a higher validation loss and training time are inter-related in the following manner. From \autoref{fig:val_loss} a high validation loss for hyper parameters obtained from the paper suggests that the model is not performing well on unseen data. This could be because the model has learnt the noise from the training data and is not performing well. To address this issue the model will require more training epoch to converge onto a solution.

Another reason we suspect this behaviour is that the hyperparameters obtained are from two research papers. From research \cite{kumar2022fine} we obtained the learning rate and the momentum for the experiment while the step size and gamma values wer obtained from \cite{goyal2022vision}. These parameters may not be compatible with each other as the \cite{kumar2022fine} did not publish the configurations for gamma and step size.

From these initial experiments we conducted we can say that the LLMs are sensible to identify hyperparameters for an experiment. However we need to further investigate this claim by performing additional experiments utilising different models and across different machine learning domains.

\label{results}

\section{Discussion}
Our study provides valuable insights into the utility of Language Models (LLMs) for hyperparameter configuration in fine-tuning tasks. However, certain observations from our experiments requires attention. Notably, we found that LLMs demonstrated the capacity to propose sensible initial hyperparameter configurations, as highlighted in \autoref{results}. Particularly significant is the trend where, within a constrained scope of epochs and trials, the hyperparameter settings suggested by LLMs yielded more favourable validation loss outcomes when compared to alternative strategies. This underscores the potential effectiveness of LLMs in enhancing hyperparameter optimisation, especially in scenarios where limited trials and epochs are available for a specific use case. Nevertheless, due to the exploratory nature of this study and resource limitations involving a single-threaded CPU, it's possible that the hyperparameters derived from the reference papers exhibited suboptimal performance. Future investigations involving larger datasets and increased trial numbers are warranted to validate these findings. By doing so, we aim to ascertain the robustness and generalizability of this phenomenon.
it is important to acknowledge several limitations and engage in a thoughtful discussion of their implications.
\subsection{Limitations}

\subsubsection{Resource Limitations:} 
Our study was conducted with resource constraints, utilising a single-threaded CPU. This might have impacted the performance and thoroughness of the experiments, potentially influencing the effectiveness of hyperparameter recommendations.

\subsubsection{Limited Exploration:} 
The experiments were carried out with a restricted number of trials and epochs due to resource constraints. This limitation could have hindered the comprehensive exploration of the hyperparameter search space, possibly affecting the generalizability of the findings.

\subsubsection{Model Specificity:} 
The outcomes of this study are particularly relevant to the specific LLM model used (ChatGPT-4) and might not be directly transferable to other LLMs or models with distinct characteristics.

\subsection{Assumption of Expertise:} 
The imitation-based prompts assume a certain level of expertise from LLMs, which might not be accurate in all situations.

\subsubsection{Hallucination and Inaccurate Information:} 
An important limitation arises from the potential for LLMs to generate responses that contain incorrect or fabricated information, a phenomenon known as "hallucination." This limitation stems from the model's training data, which might include misinformation, bias, or inaccuracies present in online content. Consequently, LLM-generated suggestions could inadvertently lead to suboptimal hyperparameter configurations that might appear sensible but are based on inaccurate premises. 

\subsection{Knowledge Cutoff and Currency: }
An inherent limitation of GPT-4 is its knowledge cutoff, which restricts its familiarity with events, developments, and information beyond a certain date. Since GPT-4's training data includes information only up to a specific point in time (e.g., September 2021), any advancements, trends, or changes in the field of interest that occurred after that date are outside the model's awareness. This limitation can impact the accuracy and relevance of LLM-generated suggestions, particularly in rapidly evolving domains where recent information is crucial for optimal hyperparameter configurations or utilising model and datasets curated after September of 2021.
\label{discussion}

\section{Conclusion}
Our study emphasises the significance of leveraging LLMs to automate and streamline the configuration process, reducing the reliance on manual intervention and human expertise. By harnessing the power of LLMs, software engineers can benefit from their adaptability, customisation potential, and ability to generate scenario-specific configurations.

However, it is important to note that while LLMs show promise in automating the configuration process, the observed variability in weight distributions necessitates further validation and refinement. In conclusion, this study contributes to the advancement of intelligent and adaptive software systems by demonstrating the effectiveness of LLMs in autonomously configuring software tools. The inherent variability of configurations within a given scenario and the variability of configurations across different scenarios highlight the adaptability and customisation potential of LLMs. This research opens up new possibilities for efficient and optimised software engineering practices, facilitating the deployment of software systems that align closely with the specific needs and requirements of diverse application domains.
\label{conclusion}

\bibliographystyle{IEEEtran}
\bibliography{references}

\end{document}